\documentclass[a4paper,10pt]{article}
\usepackage{theorem,amssymb}
\usepackage{graphicx}
\def\eps{\varepsilon}

\def\Var{\mathrm{Var}\,}
\def\Cov{\mathrm{Cov}\,}
\def\COV{\mathbf{Cov}\,}
\def\<{\langle}
\def\>{\rangle}

\begin{document} 

\title{Fading channel estimation for free-space continuous-variable secure quantum communication}
\author{L\'aszl\'o Ruppert$^{1,*}$, Christian Peuntinger$^{2,3,\dag}$, Bettina Heim$^{2,3,\ddag}$,\\ Kevin G\"unthner$^{2,3}$,
Vladyslav C. Usenko$^{1}$, Dominique Elser$^{2,3}$,\\ Gerd Leuchs$^{2,3}$, Radim Filip$^1$, and Christoph Marquardt$^{2,3}$\medskip\\ 
\small $^1$ Department of Optics, Palack\'y University, 17. listopadu 12, 771 46 Olomouc, Czech
Republic\\
\small $^2$ Max-Planck-Institut f\"ur die Physik des Lichts,Staudtstr. 2, 91058 Erlangen, Germany\\
\small $^3$ Institut f\"ur Optik, Information und Photonik, Universit\"at Erlangen-N\"urnberg, Staudtstr. 7/B2,\\ 
\small 91058 Erlangen, Germany\\
\small $\dag$ Currently at inno-spec GmbH, Sigmundstr. 220, 90431 N\"urnberg, Germany\\
\small $\ddag$ Currently at OHB System AG, Manfred-Fuchs-Str. 1, 82234 Oberpfaffenhofen, Germany \\
\small\textit{$^{*}$E-mail: ruppert@optics.upol.cz}}
\date{}


\maketitle

\begin{abstract}
We investigate estimation of fluctuating channels and its effect on security of continuous-variable quantum key distribution. We propose a novel estimation scheme which is based on the clusterization of the estimated transmittance data. We show that uncertainty about whether the transmittance is fixed or not results in a lower key rate. However, if the total number of measurements is large, one can obtain using our method a key rate similar to the non-fluctuating channel even for highly fluctuating channels. We also verify our theoretical assumptions using experimental data from an atmospheric quantum channel. Our method is therefore promising for secure quantum communication over strongly fluctuating turbulent atmospheric channels.

\end{abstract}



\section{Introduction} 

Quantum cryptography is well known to be the method for secure communication based on mathematically secure cryptosystems (such as the one-time pad \cite{Shannon1949}). Quantum key distribution (QKD) protocols \cite{Gisin2002,Scarani2009,Diamanti2016} are aimed at distributing secret keys between two trusted parties. Recent developments in the field of QKD are concerned with continuous-variable (CV) protocols \cite{Braunstein2005} based on Gaussian encoding \cite{Weedbrook2012a} of continuous observables, such as the field quadratures. The security of Gaussian CV QKD with coherent \cite{Grosshans2002} and squeezed \cite{Cerf2001} states of light was shown against collective attacks \cite{Navascues2006,Garcia2006,Pirandola2008} using the optimality of Gaussian attacks, and later extended to general attacks in the asymptotic regime \cite{Leverrier2013} as well as in the finite-size regime for certain protocols (see \cite{Diamanti2015} for review of CV QKD security proofs). Gaussian CV QKD protocols  based on coherent \cite{Jouguet2013} and squeezed states \cite{Madsen2012} were successfully tested in the channels with fixed transmittance and were also studied for free-space atmospheric channels with transmittance fluctuations \cite{Usenko2012}.

One of the major elements of CV QKD protocols is the channel estimation because it allows the trusted parties to assess the upper bound of the information leakage based on the parameters of the channel. The estimation, however, is never perfect in practice since the number of signals is limited. The issue was addressed for coherent \cite{Leverrier2010, Thearle2016}, squeezed-state \cite{Ruppert2014} protocols and even for the measurement-device-independent settings \cite{Papa2017} in fixed-type channels, i.e., channels in which the transmittance is typically stable (e.g., fiber-optical links). It was shown that for a set of fixed parameters, the channel estimation procedure can be optimized and that a double modulation can be used to approach the best performance of the protocols \cite{Ruppert2014, Chen2019}. The problem, however, becomes more complex when the transmittance of the channel fluctuates because in this case the statistics of the fluctuations must be estimated as well \cite{Usenko2012}. The fluctuating (fading) channels on the other hand are important for CV QKD because channel fading is typically observed in the atmosphere, where it is caused by air turbulence. Therefore any implementation of free-space CV QKD which does not rely on fiber-optical infrastructure has to deal with the estimation of fading channels, especially QKD implementations aiming for long-distance extraterrestrial QKD through a satellite \cite{Liao2017,Bed2017}. Importantly, the channel has to be estimated using the same quantum states as used for transmitting the key because otherwise (if bright probe pulses are used) an adversary could manipulate the signal and the bright pulses in different ways and therefore mimic the eavesdropping attacks \cite{Usenko2016}. On the other hand the estimation using quantum states with relatively low energy is far from being precise. Thus, additional methods have to be developed to improve such estimation procedures. The investigation of atmospheric channels has been a popular subject in recent years \cite{Wang2018, Chai2019, Chai2019b, Huang2019, Shen2019},but these studies typically consider a well-defined class of atmospheric transmittance (not a general transmittance distribution) and do not include all finite-size effects.

In the present paper we study the possibility of estimating fluctuating channels using quantum signals and Gaussian modulation. We analyze the estimation of fading channels theoretically and show that the imperfect estimation influences CV QKD negatively. To solve this issue, we propose the clusterization of the transmitted data, which can practically compensate for the negative impact of channel fading on CV QKD for sufficiently large data sets. We verify the results using experimental data from Gaussian modulated coherent states and their homodyne detection after passing through a fluctuating atmospheric channel. An important note is that although we used experimental data from atmospheric channels only, the given model does not rely on the specific transmittance pattern of an atmospheric channel. Our proposed estimation method works for any kind of transmittance distribution, so the results can be applied to a wide range of fading channels (actually, partially it can be even applied to fiber channels, as there would be no point in using more than one cluster due to the already small channel fluctuations).

The paper is organized as follows: Section II describes the model applied in the article and the experiment which was used to validate it; in Section III we introduce the theory of the channel parameter estimation, mainly focusing on the fluctuating transmittance and furthermore examine how well the predicted results of the model coincide with the experimental data; in Section IV we explore the limitations and basic properties of the clusterization method and in Section V we discuss the results and give our conclusions.


\section{Preliminaries}

\subsection{The model}

\begin{figure}[!ht]
\begin{center}
  \includegraphics[width=11cm]{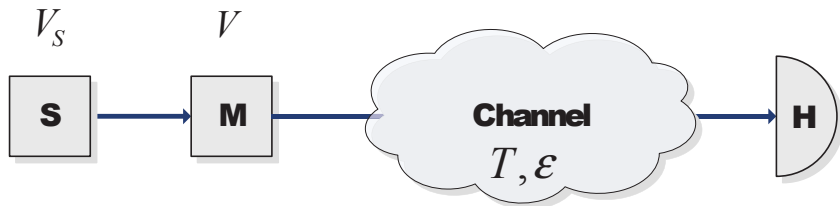}
   \caption{Prepare and measure Gaussian CV-QKD using a source ($S$) and a quadrature modulator ($M$) at Alice's side and a homodyne detector ($H$) at Bob's side. \label{method}}
 \end{center}
\end{figure}

We consider the generic CV QKD protocol, in which Alice sends CV quantum states of light to Bob through a channel that is under control of an eavesdropper, Eve. The source states are Gaussian squeezed or coherent states, Alice applies a Gaussian displacement to them using a phase and (or, in case of a squeezed-state protocol) amplitude quadrature modulator, and Bob performs homodyne detection on the output of the channel (Fig.~\ref{method}).

The preparation and detection of quantum signals is relatively fast (the channel fluctuations are on the order of several kHz, while the signal repetition rate is in the MHz-range), so for each signal sent through a fluctuating channel the transmittance can be considered stable. In this case the transfer of the quadrature variable through a lossy and noisy channel can be described in the Heisenberg picture by the following evolution (the same holds for the $p$-quadrature):
\begin{equation}\label{def_channel}
x_B=\sqrt{T}\cdot (x_S+x_M)+\sqrt{1-T} \cdot x_0+x_\eps, 
\end{equation}
where all variables are normally distributed with zero mean, except the transmittance parameter $T$. We suppose that the signal passes through a fluctuating channel (e.g., an atmospheric channel), that is, the value of the transmittance $T$ is an unknown, non-deterministic function of time. $x_B$ is the quadrature of the state Bob is measuring; $x_S$ is the quadrature of the signal state with a variance of $V_S$ (if the signal is a coherent state, then $V_S=1$); $x_M$ is the value of the Gaussian displacement of the signal with modulation $V$; $x_0$ is the vacuum noise with variance one (also referred to as shot-noise unit). Finally, $x_\eps$ is the excess noise of the channel with variance $\eps$. 

Note that in practice $x_S$, $x_0$ and $x_\eps$ are all noises (they are unknown quantities), so we can rewrite (\ref{def_channel}) in a simpler form:
\begin{equation}\label{def_ch2}
x_B=\sqrt T \cdot x_M+x_N,
\end{equation}
where $x_N$ is the aggregated noise with zero mean and variance $V_N:=1+\eps-T(1-V_S)$. Let us also note that we could have used a different parametrization (e.g., by putting $x_\eps$ within the parentheses to have a $\sqrt{T}$ factor, or by not separating Alice’s state into quantum fluctuation and modulation), we chose this approach, as the formulas are much simpler this way.

It is furthermore worth noting that even though the local estimator plays an important role during homodyne measurements, we did not include it in our analysis as this is rather a technical issue and can even be avoided. Similarly to our signal, the local oscillator is also subject to fluctuations and it also has to be estimated \cite{Zhao2018}, but its estimation is considerably easier due to its high power and the fact that we only need to know its intensity. So the normalization of the local oscillator can be performed with high precision. Moreover, the issue can be omitted entirely by using a local oscillator generated locally \cite{Qi2015}, which is not affected by the atmospheric channel.


\subsection{The experiment used for verification}

In Section \ref{est} to test and verify our theoretical results, we used the data obtained in the free-space experiment performed in Erlangen on the free-space link between the building of the Max Planck Institute for the Science of Light and the computer science building of the Friedrich-Alexander-University Erlangen-N\"urnberg (see \cite{Heim2014} for technical details, the same set-up was also used to share effective entanglement over the free-space link as reported in \cite{Rigas2006,Haeseler2008}). Two electro-optical modulators were used to achieve a two-dimensional modulation in the quadrature phase space mimicked by 192 different displaced coherent states following a pseudo-random two-dimensional Gaussian distribution. The states were repeatedly sent over the free-space channel with an effective sending rate of $2.48\cdot10^6$ states per second. At the remote side the channel transmittance was monitored using a tap-off followed by an intensity measurement, while the remaining signal was split on a symmetric beamsplitter and the $x$- and $p$-quadratures were simultaneously measured using homodyne detectors. The results of the intensity measurements of the channel transmittance (assumed to be unbiased and having standard deviation mainly as low as $10^{-4}$) were used as a benchmark to verify the estimation using Gaussian modulation of coherent states to study the possibility of improving the fading channel estimation using quantum signals.


\section{Estimation of channel parameters} \label{est}

In the following we discuss the estimation of the channel transmittance, when it is changing over time.


\subsection{Estimation of the transmittance for individual packages}

Let us divide the transmitted states during the observed time window into $m$ packages each containing the same number ($n$) of Gaussian states (that is, altogether we have $N=n\cdot m$ states). The time of transmission for a package should be so small that one can assume that the transmittance $T$ is constant for a particular package.

We suppose that the variance of the modulation ($V$) is known, that is, the channel at the $i$-th package ($i\in\{1,2,\dots,m\}$) can be parametrized using only two unknown parameters: $T_i$ and $\eps_i$. From each package Alice or Bob reveal some fraction ($r$) of their measured data to obtain an estimation of the current transmittance $\hat T_i$. Therefore, for every package the estimation is based on $r\cdot n$ Gaussian states. 

We verify the security of CV QKD by valuating the lower bound on the key rate (further simply referred to as the key rate) incorporating the finite size effects, which for fixed channel transmittance \cite{Leverrier2010} is
\begin{equation}\label{def_keyrate}
K=(1-r)\cdot \Bigg[K_{\infty}(T^{LOW},V_\eps^{UP})-\Delta\bigg([1-r]N\bigg)\Bigg],
\end{equation}
where $T^{LOW}$,$V_\eps^{UP}$ are estimates of channel parameters to evaluate the pessimistic (worst case) secure key rate, $\Delta$ is related to the errors during privacy amplification, and $K_{\infty}$ is the lower bound on the asymptotic key rate given by
$$
K_{\infty}(T,V_\eps)=\beta I(A:B)-S(B:E),
$$
where $\beta \in [0,1]$ is the reconciliation efficiency, $I(A:B)$ is the mutual information of Alice and Bob, while $S(B:E)$ is the maximal information Eve can retain about Bob's state (see more details on the Gaussian CV QKD security analysis in \cite{Usenko2016}). We consider reverse reconciliation, which is more robust against channel attenuation \cite{Grosshans2003}. It is also worth mentioning that the key rate is a monotonous function of  the channel parameters, so actually (\ref{def_keyrate}) is valid as we can obtain a confidence interval for the key rate based on the confidence intervals of the channel parameters, and the extreme value of the key rate will be taken for the extreme values of the parameters within these confidence intervals (due to monotonicity).

\begin{figure}[!t]
\begin{center}
  \includegraphics[width=9cm]{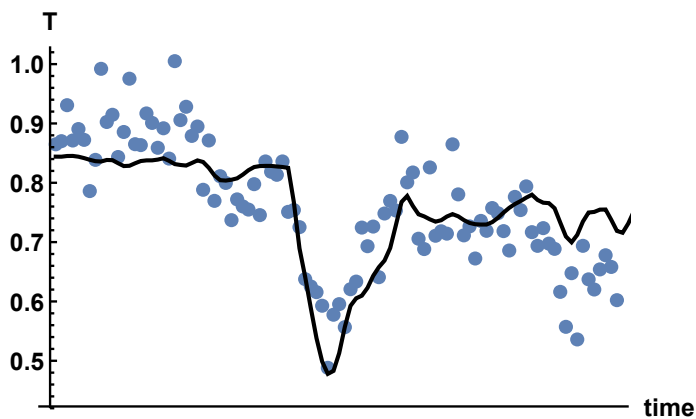}
 \caption{Fluctuations of the transmittance over the observed time (solid line). To obtain an estimation the measurement data is divided into packages such a way that the time of transmission for one package is so small that one can assume that the transmittance is constant within. From each package Alice or Bob reveal some fraction of their data to obtain an estimation of the current transmittance (blue points). \label{eta}}
 \end{center}
\end{figure}

Let us denote for the $i$-th package the realizations of $x_M$ and $x_B$ with $M_j$ and $B_j$ ($j\in \{1,2,\dots,r\cdot n\}$), respectively. We know that the covariance of $x_M$ and $x_B$ is $\Cov(x_M,x_B)=\sqrt{T_i}\cdot V=:C_{MB}$. 
That is, we can estimate the value of $\sqrt{T_i}$ (see Fig.~\ref {eta}) by 
\begin{equation}\label{est_sqrteta}
\widehat{\sqrt{T_i}}=\frac{1}{V}\cdot \widehat{C_{MB}},
\end{equation}
where we use the maximum likelihood estimator
\begin{equation}\label{est_Cov}
 \widehat{C_{MB}}=\frac{1}{r\cdot n}\sum_{j=1}^{r\cdot n} M_j B_j.
\end{equation}

\begin{figure}[!t]
\begin{center}
  \includegraphics[width=9cm]{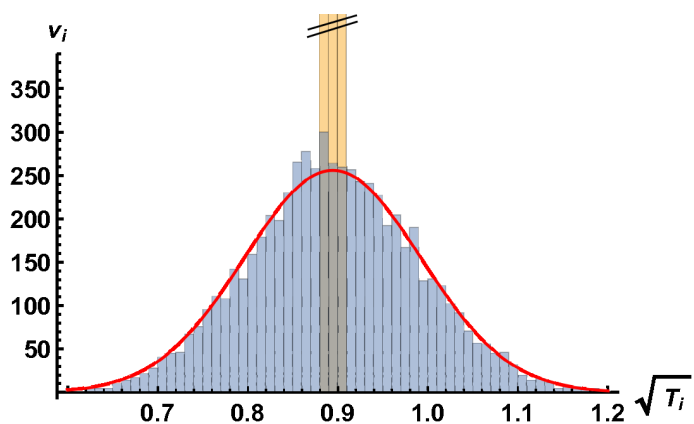}
 \caption{The histogram (with relative frequencies $\nu_i$) of the estimated square root of the transmittance values $\widehat{\sqrt{T_i}}$ (blue bars) if the real transmittance is close to 0.8 ($0.78<T_i<0.82$, that is $0.883<\sqrt{T_i}<0.906$) using the experimental data. The real transmittance distribution (yellow bars) is truncated to ensure a better view of the estimated transmittance distribution. The red solid line shows the corresponding theoretical Gaussian distribution. \label{dist}}
\end{center}
\end{figure}

It follows from the central limit theorem that $\widehat{C_{MB}}$ is approximately normally distributed, so (\ref{est_sqrteta}) is an unbiased estimator of $\sqrt{T_i}$, moreover its variance can be calculated from the parameters of the channel \cite{Ruppert2014}:
\begin{equation}\label{varsqrteta}
  \Var(\widehat{\sqrt{T_i}})=\frac{1}{r\cdot n} \cdot T_i \left(2+\frac{V_N}{T_i V}\right)
\end{equation}
Experimental data provide a very similar, approximately normal distribution (see Fig.~\ref{dist}) with a variance close to the theoretical value (0.0091 vs 0.0096) suggested by (\ref{varsqrteta}). Let us note as a reminder that in the given experiment the channel intensity was measured directly beside the quantum states to be able to compare the estimated transmittance values $\hat T_i$ from homodyne measurement (blue boxes) to their directly measured counterparts $T_i$ (yellow boxes).

Similarly, we can estimate the value of $T_i$ from
\begin{equation}\label{est_eta}
\hat{T_i}=\frac{1}{V^2}\cdot \bigg(\widehat{C_{MB}}\bigg)^2.
\end{equation}
It can be shown that $\hat{T_i}$ is an asymptotically unbiased estimation of $T_i$ with standard deviation
\begin{equation}\label{vareta}
  \Var(\hat T_i)\approx \frac{4}{r\cdot n} \cdot T_i^2 \left(2+\frac{V_N}{T_i V}\right)=:\sigma^2(T_i).
\end{equation}

Since the number of states in a package ($n$) is limited, $\hat T_i$ will be just a rough estimate of $T_i$ with a high uncertainty, which may result in a security break. A straightforward solution to overcome this problem is that instead of performing key distribution for each package independently, we merge the packages.


\subsection{The effect of fluctuations}

The covariance matrix of measurement outcomes of $x_M$ and $x_B$ is
\begin{equation}\label{covariance}
\COV(x_M,x_B)=\left(\begin{array}{cc}
V & \<\sqrt{T}\> V \\
\<\sqrt{T}\> V & \<T\> V'+\eps+1 \end{array} \right),
\end{equation}

where $\<T\>$ and $\<\sqrt{T}\>$ are the mean values of $T_i$ and $\sqrt T_i$ for the given time interval, and $V'=V+V_S-1$ (for coherent states $V'=V$).

The given covariance matrix after a fading channel can be equivalently parametrized as originating from a channel with a fixed effective transmittance and effective excess noise \cite{Usenko2012}:
\begin{equation}\label{covariance2}
\COV(x_M,x_B)=\left( \begin{array}{cc}
V & \sqrt{T_{\mathrm{eff}}} V \\
\sqrt{T_{\mathrm{eff}}} V & T_{\mathrm{eff}} V'+\eps_{\mathrm{eff}}+1 \end{array} \right),
\end{equation}

Using $\Var(\sqrt{T})=\<T\>-\<\sqrt{T}\>^2$ and the equivalence of (\ref{covariance}) and (\ref{covariance2}), one can obtain the effective transmittance as
\begin{equation}\label{effT}
T_{\mathrm{eff}}=\<\sqrt{T}\>^2,
\end{equation}
while the effective excess noise is 
\begin{equation}\label{effEps}
\eps_{\mathrm{eff}}=\eps+\Var(\sqrt{T}) V'.
\end{equation} 
We know that the key rate is most sensitive to the excess noise. Therefore, if the channel is heavily fluctuating, then $\Var(\sqrt{T})$ and the respective effective excess noise will be large, which results in a low key rate or could even lead to a security break \cite{Usenko2012}. This can be partially compensated by the use of squeezed states \cite{Derkach2018} or by stabilizing the channel \cite{Usenko2018}.


\begin{figure}[!t]
\begin{center}
  \includegraphics[width=9cm]{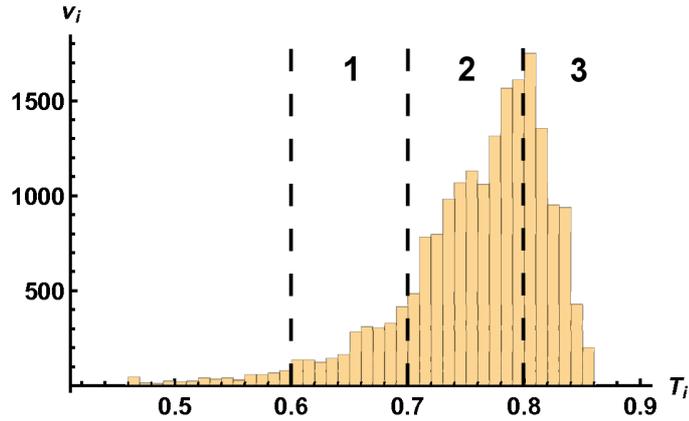}
 \caption{\label{cluster} The histogram of transmittance values $T_i$ in our experiment. The vertical lines define a possible clusterization of the data to 3 clusters: [0.6,0.7], [0.7,0.8], [0.8,0.9]. The key rate is evaluated for every cluster independently.}
 \end{center}
\end{figure}

\subsection{Data clusterization and empirical distribution}

To avoid the negative effect of a heavily fluctuating channel we do not merge every data package into a single covariance matrix. Instead we can clusterize (split into $C$ clusters) the packages (see Fig.~\ref{cluster}). The main idea is that if the fluctuation of the channel within each cluster becomes significantly lower, then according to (\ref{effEps}), the effective excess noise will decrease, which results in an increased key rate.
Since each data point is generated independently of the others, if we take a data point from a specific cluster, it will not contain any information about the data points from other clusters. So we perform security analysis for each cluster independently, and the total key rate will be the sum of the key rates determined for the individual clusters.

Note that the number of clusters is a quantity to optimize since there is an obvious trade-off: 
\begin{itemize}
\item if we use too many clusters, then the clusters will contain a small number of states, so the finite size effects will be strong (the effect of $\Delta$ in (\ref{def_keyrate}) will be dominant for fewer data points in a cluster),
\item if we do not use enough clusters, then the fluctuation within a single cluster will be very large resulting in a too large effective excess noise (see the improvement using more clusters in Fig.~\ref{rate_clusters}).
 \end{itemize}

To evaluate the key rate for a given cluster we have to estimate the transmittance statistics. This task is, however, far from trivial, since by simply averaging the estimated transmittance values belonging to this cluster we would obtain a biased estimate. The reason for this is simple: we are interested in the mean of the actual values of transmittance ($T_i$), while we can perform the clustering only based on the estimated transmittance values ($\hat{T_i}$).

\begin{figure}[!t]
\begin{center}
  \includegraphics[width=9cm]{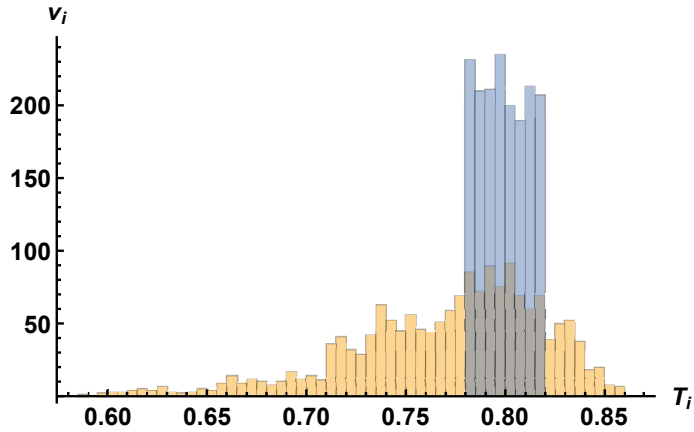}
 \caption{The histogram of estimated transmittance values $\hat{T_i}$ between 0.78 and 0.82 (blue bars) and the histogram of their actual values $T_i$ (yellow bars) using the experimental data.\label{cond_dist}}
\end{center}
\end{figure}

To highlight this effect let us look at the example of taking the estimates $\hat{T_i}$ falling into the cluster of $[0.78,0.82]$ (Fig.~\ref{cond_dist}, blue bars). If we check how the real transmittance values of these packages are distributed (yellow bars), we can see that the actual transmittance values are quite irregular and can reach well over the boundaries of the given cluster. Note that this is in some sense the opposite of Fig.~\ref{dist}, where we define the same cluster by real transmittance and investigate the estimated transmittance distribution for these states (which is much more regular).

In this particular example the average of the estimated transmittance values $\<\hat T_i\>$ belonging to this cluster is about $0.8$, while the average of their actual values $\<T_i\>$ is about $0.769$ (which is actually outside of the investigated interval). So if we evaluate the key rate simply from the average of the estimated value $\<\hat T_i\>$, we get a biased estimation, which may result in over- or underestimated key rate.


 \subsection{Confidence intervals of channel parameters}\label{conf_int}

For the evaluation of the key rate one should use the covariance matrix defined in (\ref{covariance}). But since we do not know the exact values of the parameters of the channel, we should use the worst case scenario from the appropriate confidence intervals (similarly as it was discussed for the fixed channel case below (2)). It is easy to see that in the worst case we should use the lower bound $\<\sqrt T\>^{\mathrm{LOW}}$ for $\<\sqrt T\>$, while the upper bounds $\< T\>^{\mathrm{UP}}$ and $\eps^{\mathrm{UP}}$ for $\< T\>$ and $\eps$, respectively.

\begin{figure}[!t]
\begin{center}
\includegraphics[width=9cm]{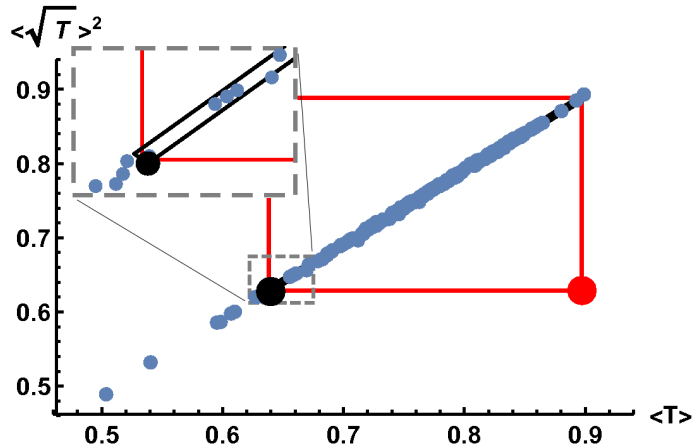}
 \caption{The value of $\<\sqrt T\>^2$ as a function of $\< T\>$ using experimental data. The two-dimensional confidence interval (using the mean plus-minus two times standard deviation) is plotted with a red rectangle, the worst case scenario (that is, which parameters provide the minimal key rate within the given confidence interval) is plotted with the big red dot. (inset) Inside the gray dashed rectangle is the magnification of the left bottom corner of the red rectangle. \label{corr}}
\end{center}
\end{figure}

However, we can see a strong correlation between the values of $\<\sqrt T\>^2$ and $\< T\>$ (see Fig.~\ref{corr}). That means that if we use rectangular confidence zones, then $\<\sqrt T\>^{\mathrm{LOW}}$, $\< T\>^{\mathrm{UP}}$ will result in a large $\Var(\sqrt{T})$ and so we will have hugely increased excess noise. 

\begin{figure}[!t]
\begin{center}
\includegraphics[width=9cm]{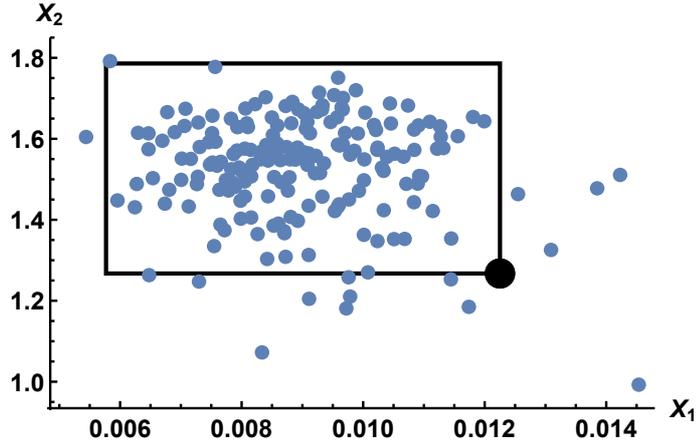}
 \caption{The value of $X_2$ as a function of $X_1$ using experimental data. The two-dimensional confidence interval (using the mean plus-minus two times standard deviation) is plotted with a black rectangle, the worst case scenario is plotted with the black dot. \label{corr2}}
\end{center}
\end{figure}

To exploit the correlation we introduce the following transformed variables 
$$
X_1:=\< T\>-\<\sqrt T\>^2=\Var(\sqrt{T})
$$ 
and 
$$
X_2:=\< T\>+\<\sqrt T\>^2,
$$ 
and we calculate rectangular confidence intervals for these variables (see Fig.~\ref{corr2}) similarly to the approach used in \cite{Ruppert2014}. This will result in the original variables in a much tighter, diagonal rectangular confidence interval (see Fig.~\ref{corr} inset, black rectangle).

It is easy to see that for the worst case scenario we should use the upper bound for $X_1^{\mathrm{UP}}$ and the lower bound for $X_2^{\mathrm{LOW}}$, so using these new variables we can obtain the following lower bound for the effective transmittance (\ref{effT})
\begin{equation}\label{estT}
T_{\mathrm{eff}}^{\mathrm{LOW}}=\frac{X_2^{\mathrm{LOW}}-X_1^{\mathrm{UP}}}{2}
\end{equation}
and the upper bound for the effective excess noise (\ref{effEps})
\begin{equation}\label{estEps}
\eps_{\mathrm{eff}}^{\mathrm{UP}}=\eps^{\mathrm{UP}}+X_1^{\mathrm{UP}}\cdot V'.
\end{equation}
Note that we detailed here only how to obtain confidence intervals for the transmittance parameters as the same for excess noise ($\eps^{\mathrm{UP}}$) is already discussed in the literature (see for example \cite{Ruppert2014}).

The main advantage of using this parametrization is that in this case we explicitly obtain an upper bound on the channel fluctuation $\Var(\sqrt{T})$. The simpler method yields $\< T\>^{\mathrm{UP}}-(\<\sqrt T\>^{\mathrm{LOW}})^2$ instead, which typically results in a 20-30 times worse estimation of $\Var(\sqrt{T})$ (for comparison see Fig.~\ref{corr}, red and black big dots). From (\ref{effEps}) we know that a much higher value of $\Var(\sqrt{T})$ results in a significantly higher effective excess noise, thus, in a lower key rate. The significance of this comes from the fact that the value of $\varepsilon$ in practical scenarios is usually not large, so actually, the biggest restriction on the key rate comes not from actual noise, but the estimation of the noise. So if we can improve the estimation of the transmittance parameters significantly, that will improve the estimation of effective excess noise significantly, thus resulting in a higher key rate.

Note that the above described method works for any given cluster. This also includes the case when there is no clusterization ($C=0$), i.e., when all data are included into the calculation.


\section{Semi-analytical investigation of the scheme} 

In the previous section we described how to estimate the channel parameters, and showed that the theory behind the estimation fits the experiment very well.

In the following we evaluate the key rate in several theoretical settings to investigate the dependence on the package size and on the clusterization of the packages.


\subsection{The package size}

First we assume that the transmittance of the fluctuating channel is normally distributed with mean $0.5$ and standard deviation $0.1$ (that is, $T_i \sim \mathcal{N} (0.5,0.1)$). Besides the transmittance distribution, we have to define seven further parameters for the numerical evaluation of specific examples. The values of the reconciliation efficiency ($\beta$), the channel excess noise ($\eps$), and the squeezing parameter of Alice's state ($V_S$) are defined in the captions. The ratio of states used for estimation ($r$) and the variance of Alice's modulation ($V$) are always optimized. The total number of states ($N$) and the number of states in each package ($n$) are in some graphs shown as variables, or denoted in the legends, or defined in the caption.

\begin{figure}[!t]
\begin{center}
  \includegraphics[width=9cm]{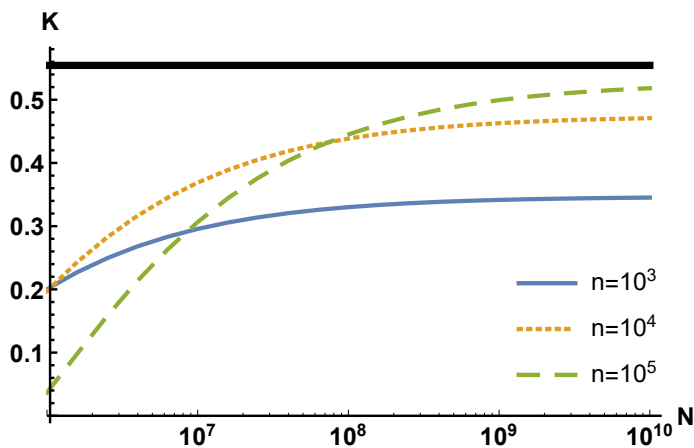}
 \caption{The key rate without clusterization as a function of the total number of states ($N=n\cdot m$) using $T_i \sim \mathcal{N} (0.5,0.1)$, $\beta=0.95, \eps=0.01, V_S=0.1$ and optimal $V$ and $r$. As a reference, the thick black line shows the asymptotically achievable key rate. \label{key_rate}}
\end{center}
\end{figure}

If we plot the key rate as a function of the total number of used states (see Fig.~\ref{key_rate}), we can see that for finite package sizes the key rate will not converge to the asymptotic value. This is due to the uncertainty of the estimators $\hat T_i$ which estimates the mean value of the transmittance for each package. The achievable key rate increases with the number of states in each package ($n$): with $n=10^5$ it will be close to the asymptotic case, however, even with $n=10^3$ it will produce a reasonable key rate. But this is true only if the total number of states ($N$) is high enough because it is also important to have a reasonable number of packages ($m$). 

So it is important to fix $n$ as high as possible to still have a (nearly) fixed value of transmittance for each package. For a fast state preparation and detection system, the value of $n$ will be higher and so we can get a better key rate, and then $N$ basically corresponds to the elapsed time. If we wait longer then the key rate will be higher, but this process saturates relatively quickly (see  Fig.~\ref{key_rate}).

As a reference, the fluctuation of an atmospheric channel is typically of the order of several kHz, while the lasers and homodyne detectors operate typically at hundreds of MHz (recently CV systems working in the GHz regime were also reported for QKD \cite{Khan2015}). The ratio between the speed of fluctuation and that of the state preparation and measurement defines the maximal size of the packages, which is of the order of the investigated values (rounding down to remain on the safe side), and then $m=10^3$ corresponds to a run which is about 1 second long.

\begin{figure}[!t]
\begin{center}
  \includegraphics[width=9cm]{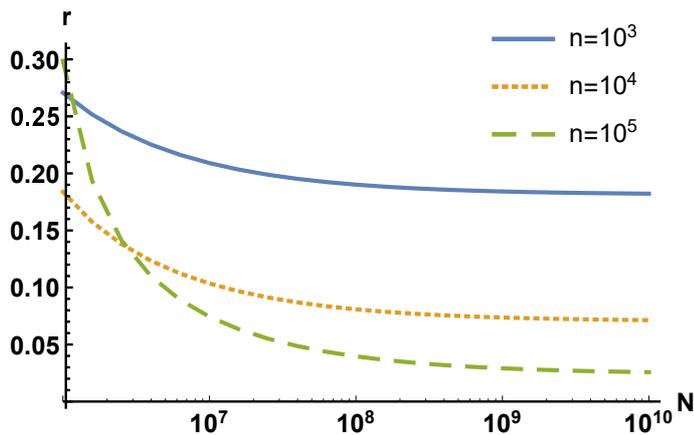}
 \caption{The optimal ratio of states used for channel estimation ($r$) without clusterization as a function of the total number of states ($N$) using $T_i \sim \mathcal{N} (0.5,0.1)$, $\beta=0.95, \eps=0.01, V_S=0.1$. \label{opt_r}}
\end{center}
\end{figure}

If we plot the optimal ratio used for estimation ($r$) (Fig.~\ref{opt_r}) we can see that if we have more states then a lower percentage is sufficient for estimation. The values saturate at a non-zero level due to the uncertain estimation for fixed size packages. Note that these values are quite low, even in the case when a package contains only $10^3$ states, 200-300 states are enough to use for channel parameter estimation, still leaving the majority of states for the key distribution itself (instead of the previously proposed 1 out of 1000 ratio \cite{Usenko2012}). This is due to the fact that even though these estimators would give a quite poor approximation of the actual value of the channel parameters, if we have many packages, the errors will cancel out. That is, even if the individual estimators are inaccurate, the effective parameters can be estimated quite well. We can also see a similar effect as in the case of the key rate: asymptotically larger package sizes result in a smaller ratio of discarded states for the channel estimation, but if the total number of states ($N$) is not so large, this can be even the opposite (i.e., smaller package size could result in a smaller optimal ratio of $r$).


\subsection{The effect of clusterization}

Now that we know how the key rate behaves without clusterization, we will investigate how much improvement can come from using clusters.

\begin{figure}[!t]
\begin{center}
  \includegraphics[width=6cm]{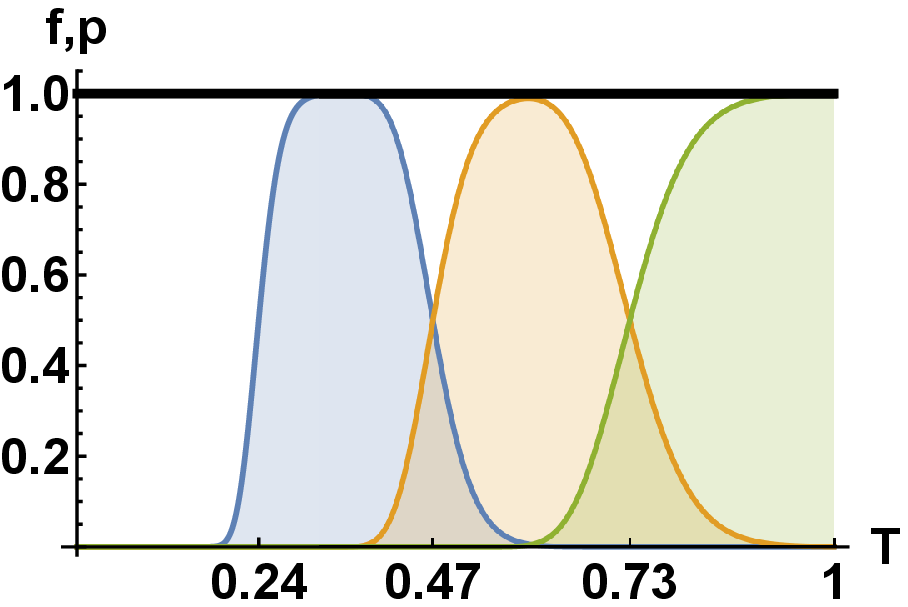}
	\includegraphics[width=6cm]{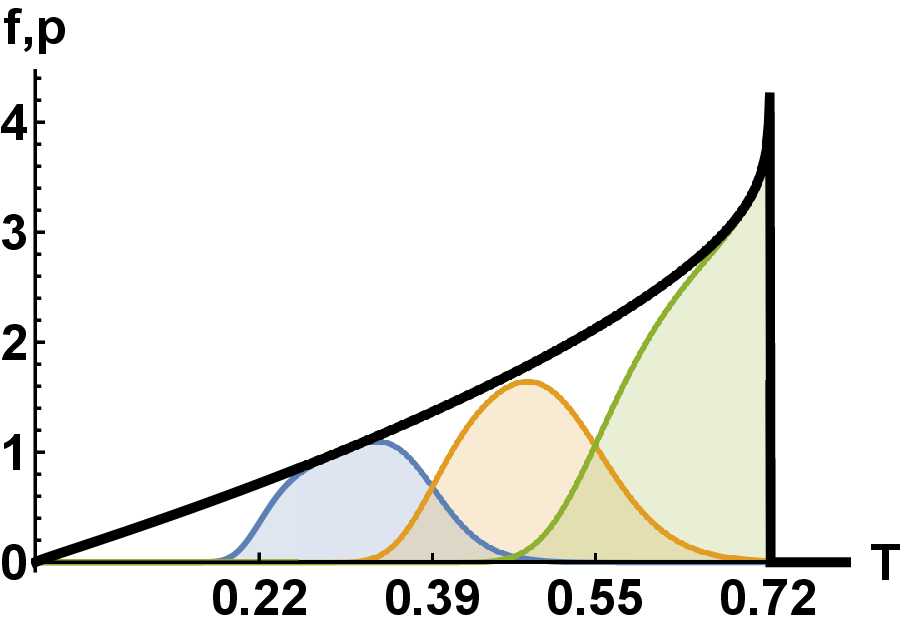}
	\includegraphics[width=6cm]{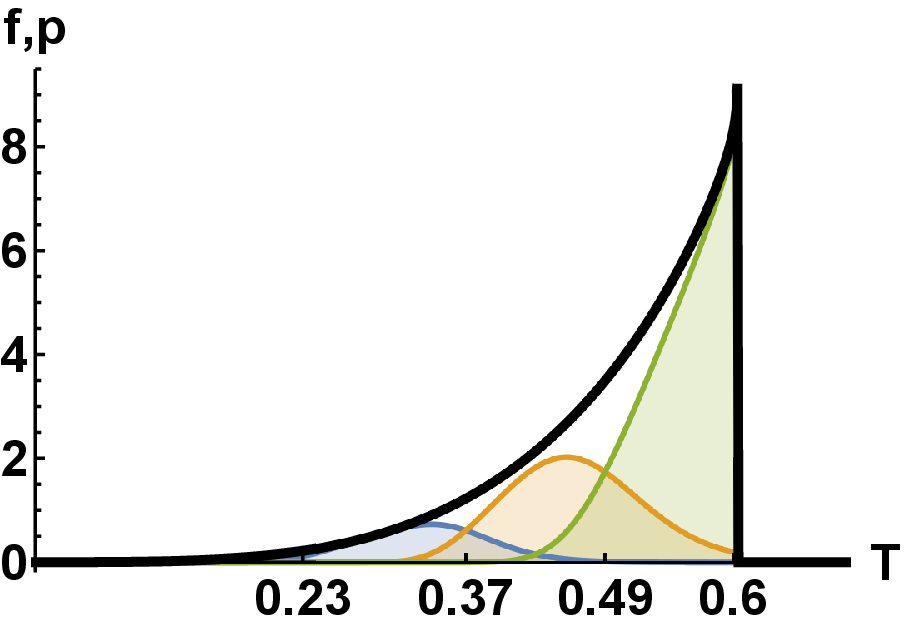}
	\includegraphics[width=6cm]{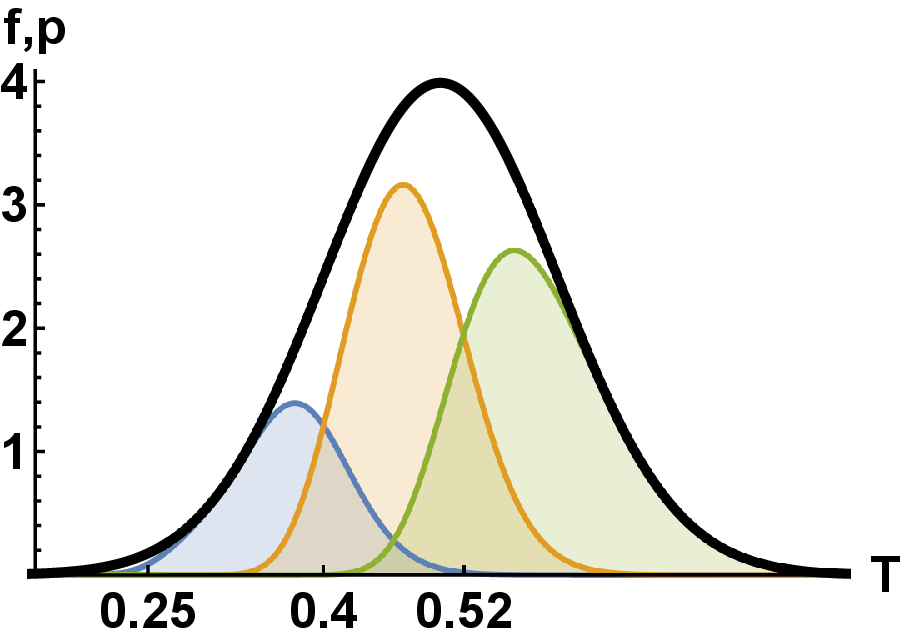}
 \caption{The optimal clusterization ($C=3$) for transmittance distribution of (top left) Uniform[0,1], (top right) Weibull[1.25,0.8], (bottom left) Weibull[1.47,0.6] and (bottom right) Normal[0.5,0.1]. The ends of cluster intervals are noted with ticks, the conditional likelihood functions $p$ of each cluster are plotted with colored areas (blue, yellow and green), the probability density function of the original distribution is plotted with black thick line. The other parameters are $\beta=0.95, \eps=0.01, V_S=0.1, n=m=10^4$. \label{clusters}}
\end{center}
\end{figure}

In the following numerical simulations we use $n=m=10^4$, so altogether a relatively large, but still experimentally feasible number of $N=10^8$ Gaussian states. We always optimize the values of $r$ and $V$ (the latter being in the order of few shot-noise units, corresponding to the vacuum quadrature fluctuations), and optimize the boundaries of the intervals defining the clusters. This optimization is easy to perform if we know the exact transmittance distribution (like in this section when using computer-generated examples), but in realistic scenarios, this process is a bit problematic as we need to know the transmittance in advance to optimize the parameters. In that case we should use a rough estimate of the expected transmittance distribution, calculate the optimal modulation $V$ and optimal ratio $r$ based on this rough estimate, send through the channel the states with the calculated modulation, reveal states from each package randomly with the calculated ratio, and then perform the proper estimation based on these revealed states (and use a clusterization if needed). Obviously, in this case, the used $V$ and $r$ for all packages will not coincide with the real optimal values, but as the key rate is not overly sensitive to these parameters, we can get very close to the optimal key rate even if the transmittance distribution used as a rough estimate is not too close to the actual distribution (that is, if there is some difference in $T_i$, there will be a smaller difference in the optimal $V$ and $r$ and an even smaller difference in the related key rates).

To investigate how the key rate depends on different types of channel fluctuations, we use four different theoretical distributions for the transmittance.
For better comparison their means are very close or equal to 0.5 ($\<T_i\> \approx 0.5$), but their variances and profiles are very different (see Fig.~\ref{clusters}, black thick lines).

\begin{figure}[!t]
\begin{center}
  \includegraphics[width=9cm]{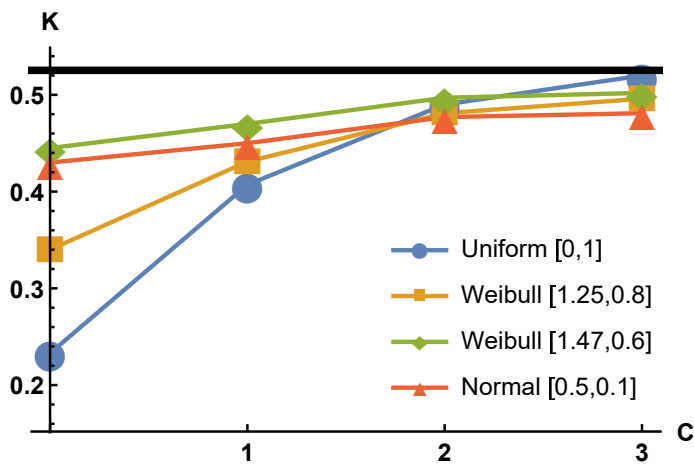}
 \caption{The optimal key rate as a function of the number of clusters ($C$), with $\beta=0.95, \eps=0.01, V_S=0.1, n=m=10^4$ for different distributions of the transmittance. In $C=0$ case we do not use any clusters, that is, we include all points into the calculation. For comparison we plotted the optimal key rate for a channel with fixed transmittance of $T=0.5$ (black thick line). \label{rate_clusters}}
\end{center}
\end{figure}

Next we need to obtain the real distribution after we select a given cluster of experimental data. This can be performed easily if we know the distribution of the transmittance. Let us assume that the aggregated likelihood function of the transmittance during the given time interval is $f(x)$. Then the probability that the actual value of the transmittance is $s$ and the estimated value is $t$ is
$$
Prob(T\approx s,\hat T \approx t) \sim f(s) \cdot \phi_s(t),
$$
where $\phi_s(x)$ is the probability density function of a normal distribution with mean $s$ and standard deviation $\sigma(s)$, which is defined in (\ref{vareta}). So the conditional distribution of the actual values supposing that the estimated value is in the interval $[T_{\min},T_{\max}]$ is given by:
$$
p(s):=Prob(T\approx s |\hat T \in [T_{\min},T_{\max}])\sim
$$
$$
\sim \int_{T_{\min}}^{T_{\max}} f(s) \cdot \phi_s(t) dt= f(s)\cdot \bigg(\Phi_s(T_{\max})-\Phi_s(T_{\min})\bigg),
$$
where heuristically the first factor describes the number of points at disposal and the second factor describes the probability of remaining in the investigated interval after adding the proper Gaussian noise coming from estimation uncertainty.
Using this conditional likelihood function $p(s)$ (see for example Fig.~\ref{clusters}, colored areas) we can obtain the mean value and the standard deviation (hence the confidence interval) of the transmittance for a given cluster (as it is discussed in Sect.~\ref{conf_int}) and for this worst case scenario we obtain the appropriate key rate. The optimized results can be seen in Fig.~\ref{rate_clusters}.

If the transmittance is completely random, that is, it is uniformly distributed between 0 and 1 (blue circles), then its fluctuation will be very high ($\Var(\sqrt{T_i})=0.055$). This case has been recently studied as a model for fast-fading channels in \cite{Papanastasiou2018}. In the asymptotic case this results in a very large effective excess noise (even up to $\eps_{eff}\approx 20\%$), therefore it is no wonder that without clustering we obtain the lowest key rate here.

We also investigated the log-negative Weibull distribution, which emerges naturally in the case of atmospheric channels as a result of beam wandering \cite{Vasylyev2012}, the main mechanism for channel fluctuations in the regime of weak atmospheric turbulence \cite{Vasylyev2016}. If we have parameters $W/a=1.25,\sigma_b=0.8$ (yellow squares) then there is a medium fluctuation ($\Var(\sqrt{T_i})=0.018$), if we use parameters $W/a=1.47,\sigma_b=0.6$ (green diamonds) we have a small fluctuation ($\Var(\sqrt{T_i})=0.0047$). Note that the given parameters are related to the beam spot to aperture size ratio, and the variance of the fluctuations of the beam around the aperture, respectively.

Finally, we plotted a normally distributed transmittance with mean $0.5$ and standard deviation $0.1$ (red triangles). The fluctuation of the transmittance in this case is also small ($\Var(\sqrt{T_i})=0.0052$), it is a bit higher than in the second Weibull case, and the related key rate is also a bit smaller.

The result is clear and not surprising, smaller fluctuations result in a higher key rate. However, if we introduce a single cluster (as it is proposed in \cite{Usenko2012}) this difference becomes smaller and we can get reasonable key rate even if the fluctuation of the transmittance is high. If we add more clusters we see the same effect: we have a higher and higher key rate, so the effect of the original fluctuation vanishes. If the fluctuation of the transmittance is small, the increment in the key rate will be small, but for large fluctations the improvement can be substantial (we can even double the key rate by using only two clusters).

An important observation is that in the three-cluster case (Fig. \ref{clusters}),  all three clusters contain a similar number of states (which is related to the sizes of colored areas). The smallest cluster is the [0.23,0.37] for the Weibull[1.47,0.6] distribution (blue area in the bottom left subfigure), but even that consist about $10\%$ of all points, meaning a block size of $10^7$ points. This is important as one has to have large block sizes to achieve high reconciliation efficiency (we used in our numerical calculations a conservative 0.95, but in practice, it depends on many factors, which are hard to quantify). This is probably not a coincident, as there are two balancing factors in our model:
\begin{itemize}
\item  If a cluster contained a very low number of points, then the term $\Delta$ in (\ref{def_keyrate}) concerning privacy amplification would be very large, possibly resulting even in a negative key rate.
\item If a cluster contained a very large number of points, then the fluctuation of the channel within ($\Var(\sqrt{T})$) would be large too, resulting in a large effective excess noise and so in a decreased key rate. 
\end{itemize}
 So in general, we can assume that the optimal clusters do not necessarily have equal sizes, but their sizes are at least of similar magnitudes and there won't be any extra small clusters.

Note that for the discussed experimental distribution we have $\Var(\sqrt{T_i})\approx 0.0015$, meaning that its fluctuation is even smaller than the given examples, so using clusterization there would not be necessary. However, our results show that even if there was a very strong fluctuation in the channel, we could still obtain a similar key rate. And this improvement is achieved only by better data processing, with no need to change the experimental setup.


\section{Summary and Conclusion}

In the current work we developed a framework to evaluate the lower bound on the key rate for CV QKD over fluctuating channels, e.g., for atmospheric channels. We divided the states into smaller packages containing the same number of states ($n$) for which the transmittance can be considered constant. We estimated the transmittance for each package independently. This only gave us a very rough estimation of the actual value of the transmittance for a single package since $n$ is small (typically $n\le 10^4$). However, if we combine the data from all these estimators we can estimate the effective transmittance and fluctuation quite accurately, which results in a high key rate. We also demonstrated that the properties of our estimation theory fit experimental data obtained from an atmospheric QKD setup very well.

In general we can obtain a higher key rate by having a source with a higher frequency, more packages (more time) and a lower fluctuation of the transmittance. But as we showed, if we clusterize the estimated transmittance values we can almost entirely eliminate the negative effect caused by the fluctuations of the transmittance even for very strong fluctuations. For a sufficient total number of states, even using only 2-3 clusters, the original difference between the key rates vanishes and we can obtain in all cases a key rate close to the one corresponding to the non-fluctuating channel. 


\medskip
\section*{Acknowledgement}
V.~C.~U. acknowledges project LTC17086 of the INTER-EXCELLENCE program of the Czech Ministry of Education and acknowledges the support of project 19-23739S of the Czech Science Foundation. L.~R. acknowledges the support of project 19-22950Y of the Czech Science Foundation and also acknowledges the support by the Development Project of the Faculty of Science, Palack\'{y} University. The research leading to these results has received funding from the H2020 European Programme under Grant Agreement 820466 CIVIQ.

\end{document}